\documentclass[twocolumn,showpacs,amssymb,floats,prb,superscriptaddress,aps,preprintnumbers]{revtex4}
\pdfoutput=1
\usepackage{graphicx}
\usepackage{dcolumn}
\usepackage{bm}
\usepackage{textcomp}
\usepackage[colorlinks=true,citecolor=blue,linkcolor=blue,filecolor=blue,urlcolor=blue,anchorcolor=blue]{hyperref}
\usepackage{color}

\begin{document}

\preprint{Phys.\ Rev.\ B, accepted}

\title{Role of tip size, orientation, and structural relaxations in\\
first-principles studies of
magnetic exchange force microscopy\\
and spin-polarized scanning tunneling microscopy}


\author{C. Lazo}
\email{clazo@physnet.uni-hamburg.de}
\affiliation{Institute of Applied
Physics and Microstructure Research Center, University of Hamburg, Jungiusstrasse 11, 20355 Hamburg,
Germany}

\author{V. Caciuc}
\affiliation{Institut f\"ur Festk\"orperforschung (IFF),
Foschungszentrum J\"ulich, 52425 J\"ulich, Germany}

\author{H. H\"olscher}
\affiliation{Institute for Microstructure Technology,
Forschungszentrum Karlsruhe, P.O. box 36 70,
76021 Karlsruhe, Germany}

\author{S. Heinze}
\affiliation{Institute of Applied Physics and Microstructure Research Center, University of Hamburg,
Jungiusstrasse 11, 20355 Hamburg, Germany}

\date{\today}

\begin{abstract}
Using first-principles calculations based on density functional
theory (DFT), we investigate the exchange interaction between a
magnetic tip and a magnetic sample which is detected in magnetic
exchange force microscopy (MExFM) and also occurs in
spin-polarized scanning tunneling microscopy (SP-STM) experiments.
As a model tip-sample system, we choose Fe tips and one monolayer
Fe on W(001) which exhibits a checkerboard antiferromagnetic
structure and has been previously studied with both SP-STM and
MExFM. We calculate the exchange forces and energies as a function
of tip-sample distance using different tip models ranging from
single Fe atoms to Fe pyramids consisting of up to 14 atoms. We
find that modelling the tip by a single Fe atom leads to
qualitatively different tip-sample interactions than using
clusters consisting of several atoms. Increasing the cluster size
changes the calculated forces quantitatively enhancing the
detectable exchange forces. Rotating the tip with respect to the
surface unit cell has only a small influence on the tip-sample
forces. Interestingly, the exchange forces on the tip atoms in the
nearest and next-nearest layers from the apex atom are
non-negligible and can be opposite to that on the apex atom for a
small tip. In addition, the apex atom interacts not only with the
surface atoms underneath but also with nearest-neighbors in the
surface. We find that structural relaxations of tip and sample due
to their interaction depend sensitively on the magnetic alignment
of the two systems. As a result the onset of significant exchange
forces is shifted towards larger tip-sample separations which
facilitates their measurement in MExFM. At small tip-sample
separations, structural relaxations of tip apex and surface atoms
can either enhance or reduce the magnetic contrast measured in
SP-STM.
\end{abstract}

\maketitle

\section{Introduction}

Recent advances in magnetic microscopy techniques
\cite{Osborne2001,Freemann2001} have allowed
fascinating new insights into magnetic properties of
nanostructures at surfaces. Such experimental measurements
challenge and drive the theoretical understanding of magnetism in
reduced dimensions and are crucial to develop new magnetic
materials. Among these experimental methods, the spin-polarized
scanning tunneling microscope (SP-STM) plays a central role as it
has opened the possibility to image magnetic structures down to
the atomic scale~\cite{2000Sci...288.1805H,%
Smith2002,2003RPPh...66..523B,2005PhRvL..94h7204K}. More recently,
the feasibility to measure even exchange forces between a magnetic
tip and a magnetic sample directly has been demonstrated using an
atomic force microscope equipped with a magnetically coated
tip.~\cite{2007Natur.446..522K} This new technique, denoted as
magnetic exchange force microscopy (MExFM) \cite{Wiesendanger1990}
opens new vistas in atomic-scale magnetic imaging~\cite{footnote_MFM}
as it is applicable to all magnetic surfaces, i.e.~conducting as
well as insulating systems,\cite{2007Natur.446..522K} e.g.~magnetic
molecules. However, as in SP-STM the interpretation of measurements
by MExFM is not straightforward and the development of theoretical
models and tools to understand them is essential.

Scanning probe-microscopy techniques are capable to
operate at the atomic scale and, if they are spin sensitive, down
to the single spin level. Thus, quantum mechanical effects on the
microscopic level are crucial and need to be properly taken into
account to achieve a theoretical understanding of these
techniques. In the past, first-principles calculations based on
density functional theory (DFT) have demonstrated their great
potential in this respect. In fact, they have become indispensable
and versatile tools to study real nanostructures in order to gain
a qualitative and often even quantitative understanding. However,
one is frequently limited by the size of the system which can be
considered and the level of approximation which is used.

In the case of STM, the Tersoff-Hamann
model\cite{PhysRevLett.50.1998} and its generalization to
spin-polarized tunneling~\cite{PhysRevLett.86.4132} is most often
used to calculate and interpret experimental images. However, the
interactions with the tip are neglected in this model. Therefore,
it is only necessary to calculate the electronic and magnetic
structure of the isolated sample, in particular, the local density
of states at a few {\AA}ngstr\"oms above the sample surface.
Naturally, the approximation of the Tersoff--Hamann model breaks
down at small tip-sample distances. Effects of tip-sample
interaction on the tunneling current have been investigated
theoretically in the past concerning conventional
STM,~\cite{DiVentra1999,Hofer2001,PhysRevB.70.085405} but to our
knowledge not with respect to SP-STM.

On the other hand, to model atomic force microscopy experiments,
it is essential to calculate the forces between the sample and the
tip. For this purpose one has to include, besides the sample, some
kind of tip model in the calculations. This fact makes the
simulation of AFM experiments much more challenging, in
particular, if one allows for structural relaxations of tip and
sample which can often be crucial. Such realistic theoretical
modelling of the interaction between tip and sample, and even of
the entire experimental procedure, has become an integral and
essential part of many AFM
experiments.\cite{Perez1997, Perez1998, 2002PhRvL..88d6106S,2002ApSS..188..306F,RMP75_949,2004Nanot..15S..60F,2006PhRvB..73x5435W,Caciuc06,sugimoto1134991,Atodiresei08,Caciuc2008}
Nevertheless, in the case of MExFM, there
have been only few studies in the past.\cite{PhysRevB.52.7352,PhysRevB.56.3218,nakamura99,Foster01,2003JPSJ...72..588M,2005SurSc.590...42M}

The first theoretical study of MExFM was based on a semi-empirical
tight-binding calculation.\cite{PhysRevB.52.7352} It was shown
that the exchange forces between an iron tip and a chromium or a
nickel surface are well below 1~nN but they should be detectable
with an atomic force microscope. The authors found that the exact
morphology of the tip does not play an important role on the
results. In this work, however, relaxation of the apex atom and
the sample were neglected, and only the $d$-electrons of the
system were considered. Subsequently, Nakamura \emph{et
al}.\cite{PhysRevB.56.3218, nakamura99} employed a more
sophisticated approach based on DFT to calculate the magnetic
exchange force between two Fe(001) surfaces. Forces of a few nN
were obtained at a distance of 3 {\AA}. Additionally, the forces
exhibited an oscillatory RKKY-interaction-like behavior as a
function of separation. Even above 4\,{\AA}, the measured forces
were still within the experimental resolution limit of AFM. Later, Foster \emph{et
al}.\cite{Foster01} used a periodic unrestricted Hartree-Fock method to calculate the interaction of a spin-polarized H or Na atom with the antiferromagnetic NiO(001) surface. They found that the difference in force
over Ni atoms with opposite spins should be detectable with the AFM
for a tip-sample distance smaller than 4 \AA\ or for imaging
close to the repulsive regime. However, at such short distances,
the chemical forces can become strong and it
was speculated that instabilities may become apparent.

A more recent first-principles
study\cite{2003JPSJ...72..588M,2005SurSc.590...42M} of the
exchange force between a single iron atom, representing the tip,
and the (001) surface of the antiferromagnetic insulator NiO has
been carried out within the framework of DFT. The calculated MExFM
images show a magnetic contrast on the atomic scale when the
single Fe atom tip approaches the surface within 1\,{\AA} above
the contact point. Therefore, this work predicted the possibility
of using AFM for magnetic imaging with atomic resolution. However,
this study did not address the role of structural relaxations and
the adequacy of the single atom tip model was not investigated.

Nonetheless, these early theoretical predictions and the outlook to directly measure magnetic exchange forces encouraged  many
experimental attempts to demonstrate MExFM, focussing especially
on the (001) surface of the antiferromagnetic insulator
NiO.~\cite{Hosoi2000,Hosoi2004,Holscher2002,Langkat2003,Schmid2008}
However, it took several years before the first successful
experiment demonstrating the predicted effect was
reported.~\cite{2007Natur.446..522K,kaiser08}

Here, we apply density functional theory using the full-potential
linearized augmented plane wave (FLAPW) method to study the
interaction of a magnetic tip and a magnetic sample as it occurs
in SP-STM or MExFM. We consider one monolayer Fe on W(001) as a
model sample system which exhibits a c$(2\times2)$
antiferromagnetic structure and has been experimentally resolved
by both SP-STM~\cite{2005PhRvL..94h7204K} and
MExFM.\cite{Schmidt2008} The iron tip is modeled by a single Fe
atom as well as by Fe clusters of different size and structural
relaxations of both tip and sample due to their mutual interaction
are also included. Our results show that the relaxations depend
sensitively on the magnetic configuration between tip and sample,
i.e.~whether the tip magnetization is parallel or antiparallel to
the Fe atom below. We calculate the exchange forces and
demonstrate that their measurement in MExFM for this tip/sample
system is facilitated due to relaxations as their onset is shifted
to larger tip-sample separations. By simulating MExFM images, one
can explain the contrasts observed in recent experiments and show
that they are due to a competition between chemical and magnetic
forces.~\cite{Schmidt2008}

Concerning SP-STM, we estimate the effect of tip-sample
relaxations on the experimental corrugation amplitude, i.e.~the
maximum vertical tip height change while scanning the surface in
the constant-current mode. We find that at a tip-sample separation
of 4~{\AA} the corrugation amplitude due to relaxations is of
similar magnitude as the contribution from the spin-polarized
tunneling current. This corrugation amplitude due to the exchange
forces can either enhance or reduce the total magnetic signal as
the sign of the spin-polarized tunneling current depends on the
electronic structure of tip and sample at the Fermi energy, while
the forces depend on the total, i.e.~energy integrated,
magnetization densities.

This paper is structured as follows. In section \ref{sec:details},
we provide details on the computational method and setup of the
calculations, e.g.~the different geometries of the considered Fe
tips are presented. In section \ref{sec:results} we review the
results. First, we discuss the obtained forces on a pyramid of
five Fe atoms without structural relaxations due to tip-sample
interaction. We then consider the effect of rotating the tip and
analyze the contributions of total force originating from
different tip atoms. From these force curves we expect
considerable structural relaxations of tip and sample due to their
mutual interaction. In section \ref{sec:relaxations}, the
relaxations are shown to depend on the local magnetic
configuration between tip and sample magnetization and we find
that the onset of significant exchange forces are shifted to
larger tip-sample distances. In section \ref{sec:interaction}, we
analyze the tip-sample interaction in terms of magnetic moments
and charge density difference plots which clearly indicate that
there is an interaction of the tip apex atom with nearest and
next-nearest Fe surface atoms. In section \ref{Cmodels}, we
compare the exchange forces obtained with five Fe atoms tip to
calculations using a single Fe atom or a fourteen Fe atoms tip.
Qualitatively, the exchange interaction is similar for the two
cluster tips while the single Fe atom tip seems an inappropriate
tip model. Finally, we estimate in section \ref{sec:spstm} the
influence of tip and sample relaxations on the corrugation
amplitude obtained in the constant-current mode of SP-STM.

\section{Computational method}
\label{sec:details}

In order to gain insight into the magnetic interactions which
occur in an SP-STM or MExFM experiment between an Fe tip and a
monolayer of iron atoms on W(001), we have performed first-principles
calculations based on density functional theory within the
generalized gradient approximation (GGA)
\cite{PhysRevLett.77.3865} to the exchange-correlation potential.
We apply the full-potential linearized augmented plane wave method
as implemented in the {\footnotesize{WIEN2K}}
\cite{PhysRevB.64.195134} code.

\begin{figure}
{\includegraphics[width=0.45\textwidth]{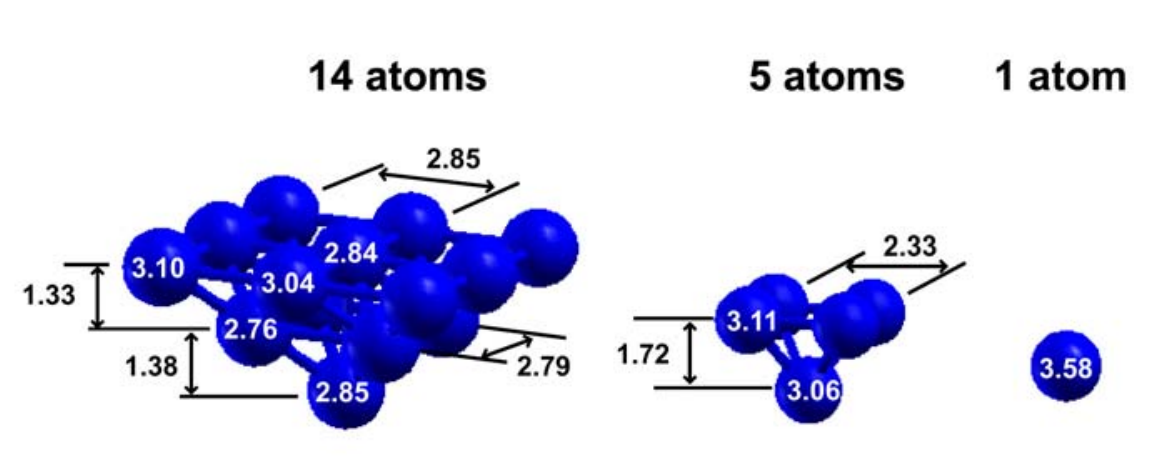}}
\caption{\label{models} (Color online) Different cluster
geometries used to model the iron tip. The magnetic moments of the
Fe atoms integrated over the muffin-tin spheres of radius $R_{\rm
MT}=2.15$~a.u.~are indicated in white in units of the Bohr
magneton, $\mu_B$. Additionally, the interlayer distances are
indicated in black in units of {\AA}ngstr\"om. The values
correspond to relaxed geometries of the isolated tip, i.e.~without
any tip-sample interactions.}
\end{figure}

We used ferromagnetic Fe pyramids in bcc-(001) orientation of
different size ranging from one to fourteen atoms to model the tip
as shown in Fig.~\ref{models}. The five-atoms tip has been fully
relaxed, i.e.~also the in-plane separation between the base atoms,
while for the fourteen atom tip only the apex atom and the four
atoms of the adjacent layer have been relaxed. The in-plane
interatomic distance between the base atoms of the fourteen atom
tip has been kept fixed at the calculated GGA lattice constant of
Fe (2.85 \AA). As can be seen in Fig.~\ref{models}, the magnetic
moments of the Fe apex atom is reduced significantly with
increasing tip size from a single atom, 3.58~$\mu_B$, to a
fourteen atom tip, 2.85~$\mu_B$, which is very similar to the
moment of 2.79~$\mu_B$ we obtained for a single Fe atom adsorbed
on the (001) surface of Fe.

\begin{figure}
{\includegraphics[width=0.48\textwidth]{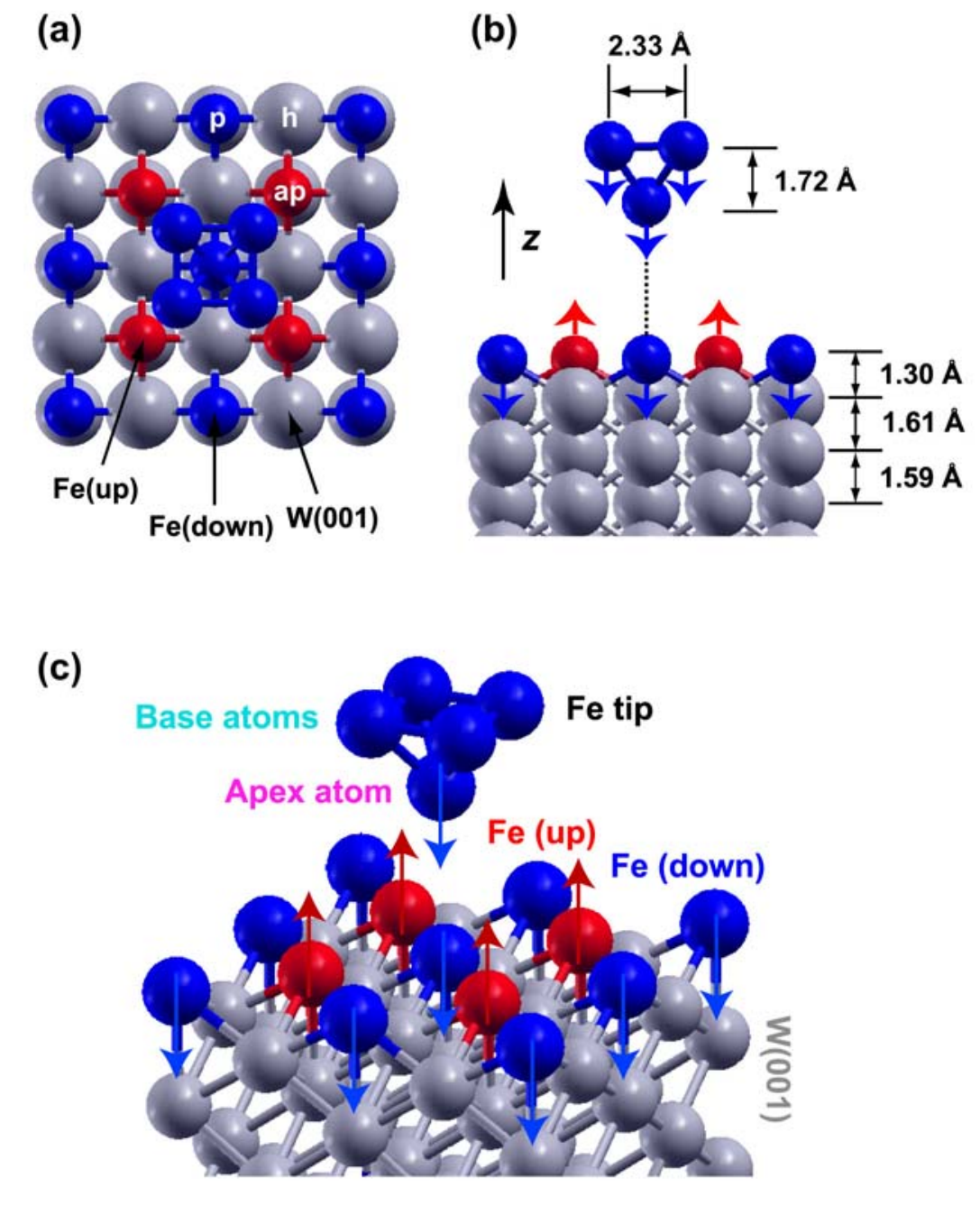}}
\caption{\label{fig2} (Color online) (a) Top view of the
c$(4\times 4)$ unit cell used to calculate the forces between the
five Fe atoms pyramid and the Fe monolayer on W(001). Sites with
parallel (p-site) and antiparallel (ap-site) alignment between tip
and surface Fe magnetic moments (indicated by arrows) are marked
as well as the hollow-site (h-site). Distances given in the side
view (b) are obtained after relaxing tip and sample independently.
$z$ is defined as tip-sample distance along the approach
trajectory (dotted line) before considering relaxations due to
tip-sample interactions. (c) 3D view of the unit cell.}
\end{figure}

The coupled system of tip and sample was calculated in a supercell
geometry, as shown in Fig.~\ref{fig2} for the example of the
five-atoms Fe tip. The monolayer of Fe on W(001) was modelled by a
symmetric slab with 5 layers of W-atoms and one layer Fe-atoms on
each side. We used the GGA lattice constant of W (3.181~\AA) which
is only 0.5\% larger than the experimental value (3.165~\AA). The
muffin-tin radii of Fe and W are 2.15 and 2.50~a.u., respectively.
The energy cut-off for the plane wave representation in the
interstitial region is $E_{\text{max}}^{\text{wf}}=11$ Ry and a
$(3\times 3\times 1)$ Monkhorst-Pack grid was used for the
Brillouin zone integration. Tip and surface were initially relaxed
independently before considering the coupled system, i.e., the
tip-sample interaction.

In two dimensions (2D) our supercell corresponds to a c$(4\times
4)$ unit cell with respect to the Fe/W(001) surface, as shown in
Fig.\,\ref{fig2}(a) for the example of the five-atoms tip. This
choice guarantees that the tip interaction with its lateral image
is negligible. E.g.~the lateral distances between adjacent
five-atoms tips are 9.0~\AA\ for the apex atom and 6.7~\AA\ for
the base atoms of the tip. Our supercell is periodic also in
$z$-direction. Choosing a very large vacuum separation of 21~\AA\
between adjacent surfaces, however, allows the tip to approach the
surface without interacting with its periodic image.

The separation $z$ in the unrelaxed coupled system is defined as
the distance between the centers of the tip apex atom and surface
atom underneath. For the relaxed system the separation $z$ is
defined as the distance between the center of the tip apex and
surfaces atoms for the case of zero relaxation. As it turned out,
the relaxation of these atoms are relatively small. Therefore,
treating $z$ as a nominal distance (as if there were no
relaxations at all) does not modify the force-distance curves
much. In fact, it is reminiscent of the experimental situation
where the exact measurement of the distance between the tip apex
and surface atom is impossible, and one uses some reference
distance.

Force curves are calculated on two high symmetry points of the
surface, which are magnetically different with respect to the
magnetization direction of the iron tip pyramid: on-top of an Fe
atom with parallel magnetic moment, $F_{\rm p}(z)$ (p-site) and on
top of an Fe atom with antiparallel magnetic moment, $F_{\rm
ap}(z)$ (ap-site).

In the case of the five-atoms tip, we also investigated the effect
of rotating the tip by $45^\circ$ with respect to the $z$-axis and
the influence of structural relaxations of tip and sample due to
their mutual interaction. Upon approaching the tip to the surface
along the $z$-direction indicated by a dotted line in
Fig.\,\ref{fig2}(b), we allowed all Fe atoms of the monolayer, the
first layer of W atoms, and the Fe apex atom to relax at every
tip-sample distance $z$. The remaining $z$-components of the
forces acting on the base atoms constitute the total force on the
tip.

\section{Results}
\label{sec:results}

In the following we present the results of our first-principles
simulations of magnetic exchange force microscopy on the Fe
monolayer on W(001). In sections \ref{sec:unrelaxed} to
\ref{sec:interaction}, we focus on the five-atoms Fe tip comparing
calculations without and with structural relaxations and analyze
the electronic structure changes due to the interaction. These
results are compared in section \ref{Cmodels} with calculations
for a single Fe atom tip and a fourteen-atoms Fe tip. Finally, we
discuss the implications of tip-sample interactions on
spin-polarized STM measurements in Sec.\,\ref{sec:spstm}.

\subsection{Unrelaxed tip and sample}
\label{sec:unrelaxed}

\begin{figure}
{\includegraphics[width=0.42\textwidth]{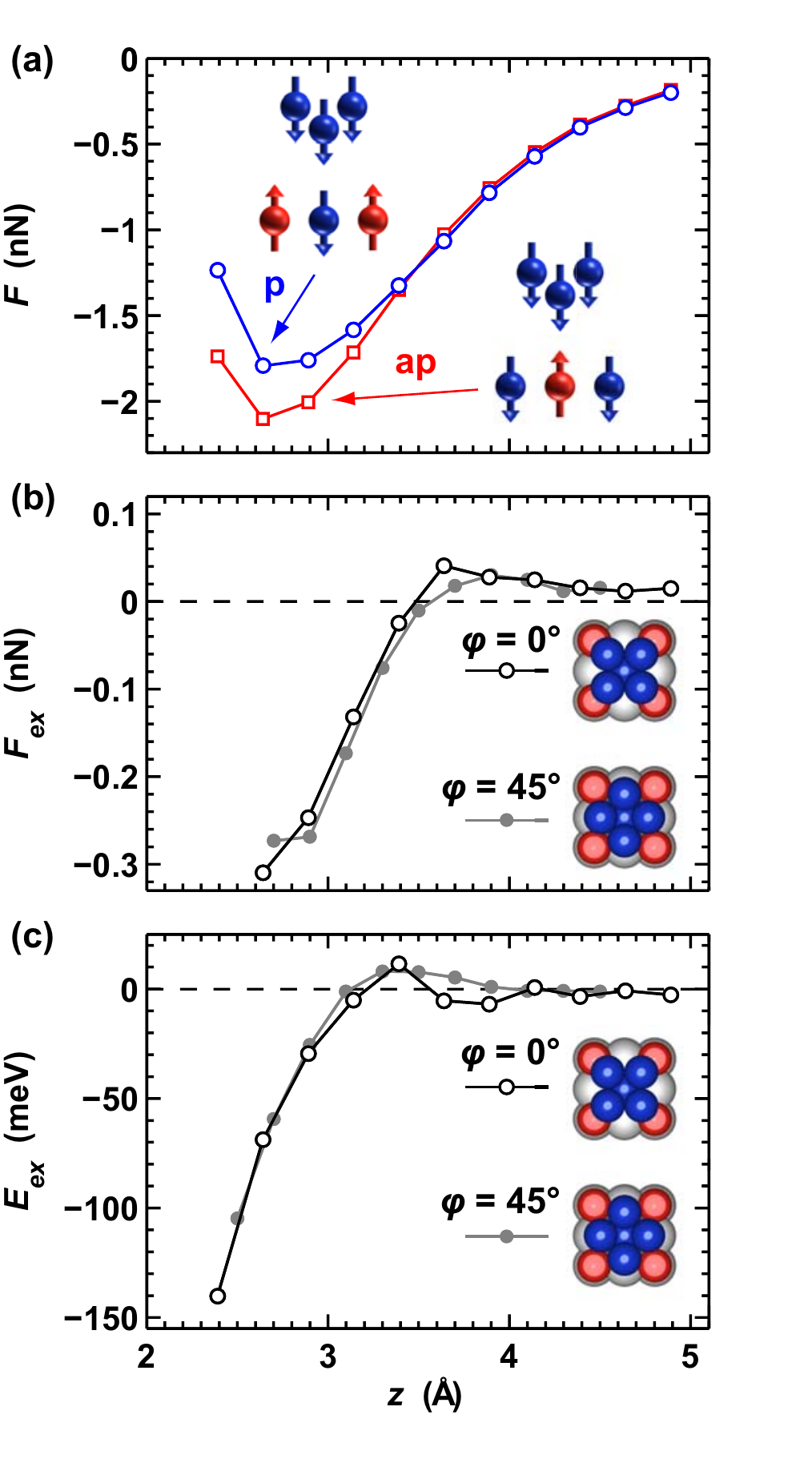}}
\caption{\label{force} (Color online) (a) Calculated force curves
for the five-atoms Fe tip as it is approached to the Fe ML on
W(001) on the ap-site, $F_{\text{ap}}(z)$, and on the p-site,
$F_{\text{p}}(z)$, of the surface, c.f.~Fig.~\ref{fig2}, for the
tip-sample system neglecting relaxations due to the interaction.
(b) The resulting exchange force,
$F_{\text{ex}}(z)=F_{\text{ap}}(z)-F_{\text{p}}(z)$, and (c) the
exchange energy
$E_{\text{ex}}(z)=E_{\text{ap}}(z)-E_{\text{p}}(z)$, as a function
of the separation, $z$, between the tip apex atom and the
approached Fe surface atom. The results for a tip rotated by
$45^{\circ}$ are plotted by filled symbols.}
\end{figure}

First, we performed separate structural relaxations of tip and
sample (c.f.\ Fig.~\ref{fig2}(b)). Then the tip was approached
vertically to the surface of the sample on the p- and ap-site
(c.f.\,Fig.\,\ref{fig2}(a)) keeping the internal geometry of the
tip and sample fixed, i.e.~neglecting structural relaxations due
to the mutual interaction. The calculated total forces acting on
the five-atoms Fe tip are shown in Fig.~\ref{force}(a). They
display an attractive interaction for the ap- and p-alignment up
to a maximum force of approximately $-2.1$ and $-1.8$~nN,
respectively, at about 2.7~\AA. The difference between the force
curves on the p- and ap-site, clearly visible in
Fig.~\ref{force}(a), is the magnetic exchange force (MExF),
$F_{\text{ex}}(z)$, defined as
\begin{equation}
F_{\text{ex}}(z)=F_{\text{ap}}(z)-F_{\text{p}}(z) \label{eq:MExF}
\end{equation}
which is depicted in Fig.~\ref{force}(b). Interestingly, the
exchange force changes its sign upon approaching the surface and
reaches significant values on the order of 0.2~nN at about
3~{\AA}. The negative sign of the exchange force indicates a more
attractive interaction for an antiparallel alignment of the
magnetization of the tip and the Fe surface atom which is being
approached (ap-site). The magnetic interaction between tip and
sample can be inferred more directly from the magnetic exchange
energy given by
\begin{equation}
E_{\text{ex}}(z)=E_{\text{ap}}(z)-E_{\text{p}}(z)
\end{equation}
and displayed in Fig.~\ref{force}(c). At large tip sample
distances, the exchange energy is very small and positive, while
it becomes quite large and negative at small separations. Figures
\ref{force}(b) and (c) include also the result of
$F_{\text{ex}}$(z) and $E_{\text{ex}}(z)$ for the cluster tip
rotated by $45^{\circ}$ with respect to the horizontal (see
Fig.~\ref{fig2}). The results show that the magnetic exchange
force and energy are hardly affected by rotating the cluster tip.

A negative sign of the exchange energy reveals that antiparallel
alignment of tip and sample magnetization, i.e.~antiferromagnetic
coupling, is energetically more favorable. This result may seem
rather surprising at first glance as one would intuitively expect
ferromagnetic coupling between the interacting Fe atoms of tip and
sample. However, as we will show in section \ref{sec:interaction}
the Fe apex atom interacts not only with the Fe surface atom
beneath it but also with the four nearest Fe neighbors of this
surface atom, c.f.~Fig.~\ref{fig2}(c). Since the magnetic moments
of the Fe atoms on the W(001) surface form an antiferromagnetic
checkerboard structure, on the ap-site the magnetization of the
tip apex atom is aligned antiparallel to the moment of the Fe
surface atom beneath it and parallel to the moments of the four
nearest neighbor Fe surface atoms and vice versa on the p-site,
c.f.~Fig.~\ref{fig2}. Therefore, if we assume ferromagnetic
coupling between individual Fe atoms there is a competition of
exchange interactions with the surface Fe atom and its nearest
neighbors.

\begin{figure}
{\includegraphics[width=0.42\textwidth]{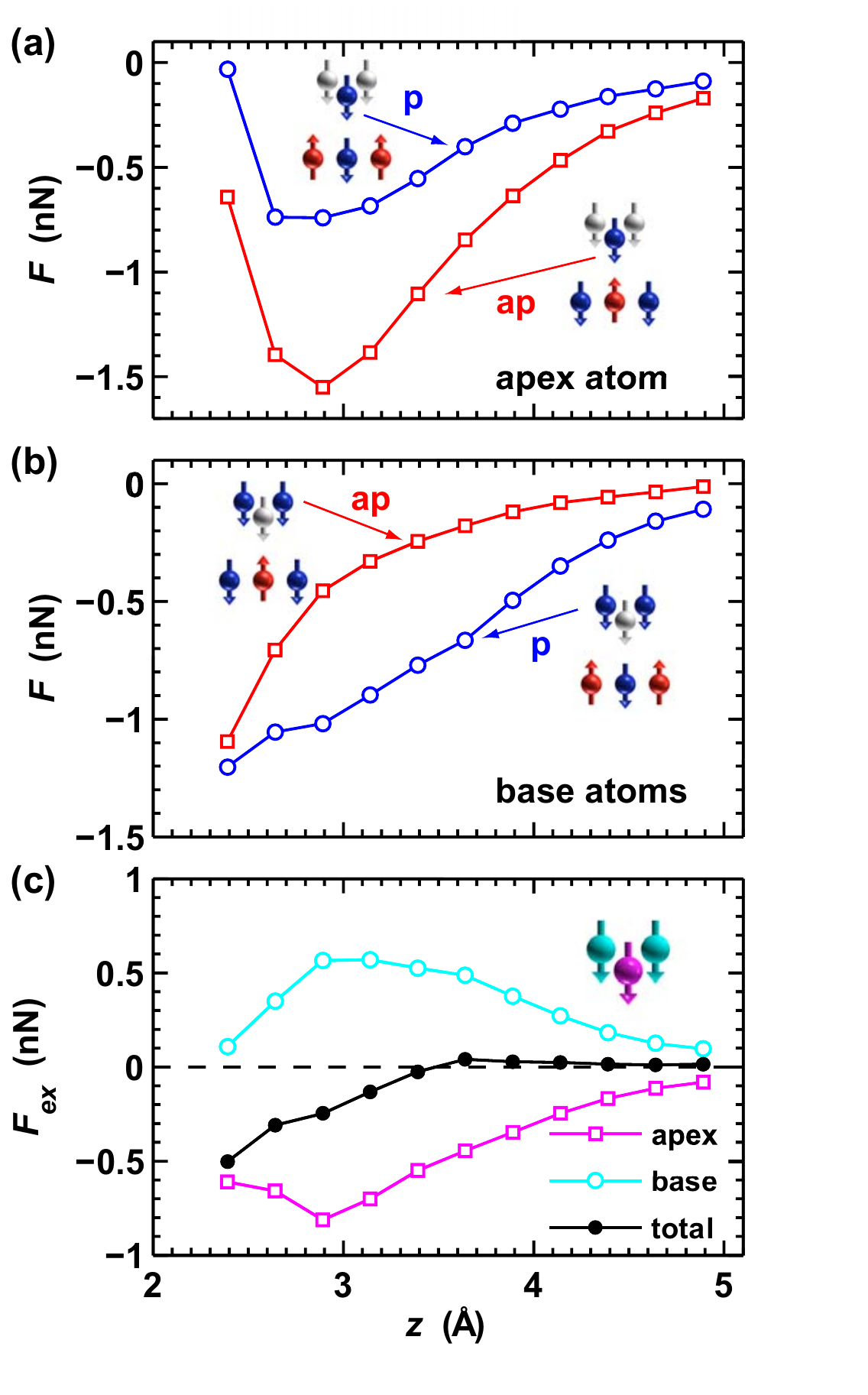}}
\caption{(Color online) \label{ForcesDecom} Decomposition of
forces acting on the unrelaxed tip for the ap- and p-
configurations. Forces acting on (a) the tip apex atom, on (b) the
base of the cluster tip, and (c) the corresponding exchange
forces.}
\end{figure}

Further insight into the tip-sample interaction and the forces
acting in the system can be obtained by decomposing the total
force on the tip. The total force acting on the cluster tip, shown
in Fig.~\ref{force}(a), is the sum of the $z$-components of the
forces acting on the tip apex atom and the four tip base atoms
which are displayed in Figs.~\ref{ForcesDecom}(a) and (b),
respectively. The apex and base tip atoms are depicted in
Fig.~\ref{fig2}. The forces acting on the tip apex atom are
qualitatively very similar to the total forces acting on the
entire cluster. However, the splitting between the forces on the
ap- and p-site is dramatically enhanced. Consequently, these large
force differences due to the magnetic interaction result in site
dependent relaxations of the tip apex atom - an effect we will
study in the next section.

The forces summed over all four base atoms, shown in
Fig.~\ref{ForcesDecom}(b), look somewhat different than those
acting on the apex atom. They can be understood if we take into
account that we sum over four atoms and that the base atoms are
1.72~{\AA} farther from the surface atoms than the apex atom,
c.f.~Fig.~\ref{models}, and consequently do not come as close to
the surface. Therefore, the force curves of the base atoms should
resemble only the part of the apex force curves at large
distances, i.e.~only the part of Fig.~\ref{ForcesDecom}(a) up to
about $z=4$~{\AA}, which is well fulfilled. Surprisingly, however,
the forces on the base atoms are larger on the p-site than on the
ap-site and therefore opposite to those on the apex atom.

The difference between the single atom forces on the p- and
ap-site can be interpreted as the individual exchange forces
acting on the apex tip and base atoms, e.g.
\begin{equation}
F_{\rm ex}^{\rm apex} (z) = F_{\rm ap}^{\rm apex}(z) - F_{\rm
p}^{\rm apex}(z).
\end{equation}
As seen in Fig.~\ref{ForcesDecom}(c), they have opposite sign and
are considerably larger than their sum, i.e.~total exchange force
acting on the tip, c.f.~Fig.~\ref{force}(c). The partial
compensation of the exchange force on the tip apex and the four
base atoms leads to a significant reduction of the total exchange
force. In addition, the exchange force on the apex atom sets in
already at much larger tip-sample distances and increasing the
contribution from the apex atom would greatly enhance the
measurable magnetic signal. This result reveals the influence of
the interaction of the sample with the tip base atoms. A realistic
model of the tip should therefore include not only a single tip
apex atom but at least some tip base atoms. In section
\ref{Cmodels}, we will explore this aspect in more detail.

The shape of the exchange force curve for the Fe apex atom
displayed in Fig.~\ref{ForcesDecom}(c) also hints at competing
exchange interactions of different sign. At large tip-sample
separations, the force is negative and rises in magnitude with
decreasing distance and reaches a local maximum of its absolute
value at $z=2.9$~{\AA} before the magnitude decreases again. A
similar shape is visible for the base atoms but with opposite sign
of the exchange force.

\begin{figure}
{\includegraphics[width=0.42\textwidth]{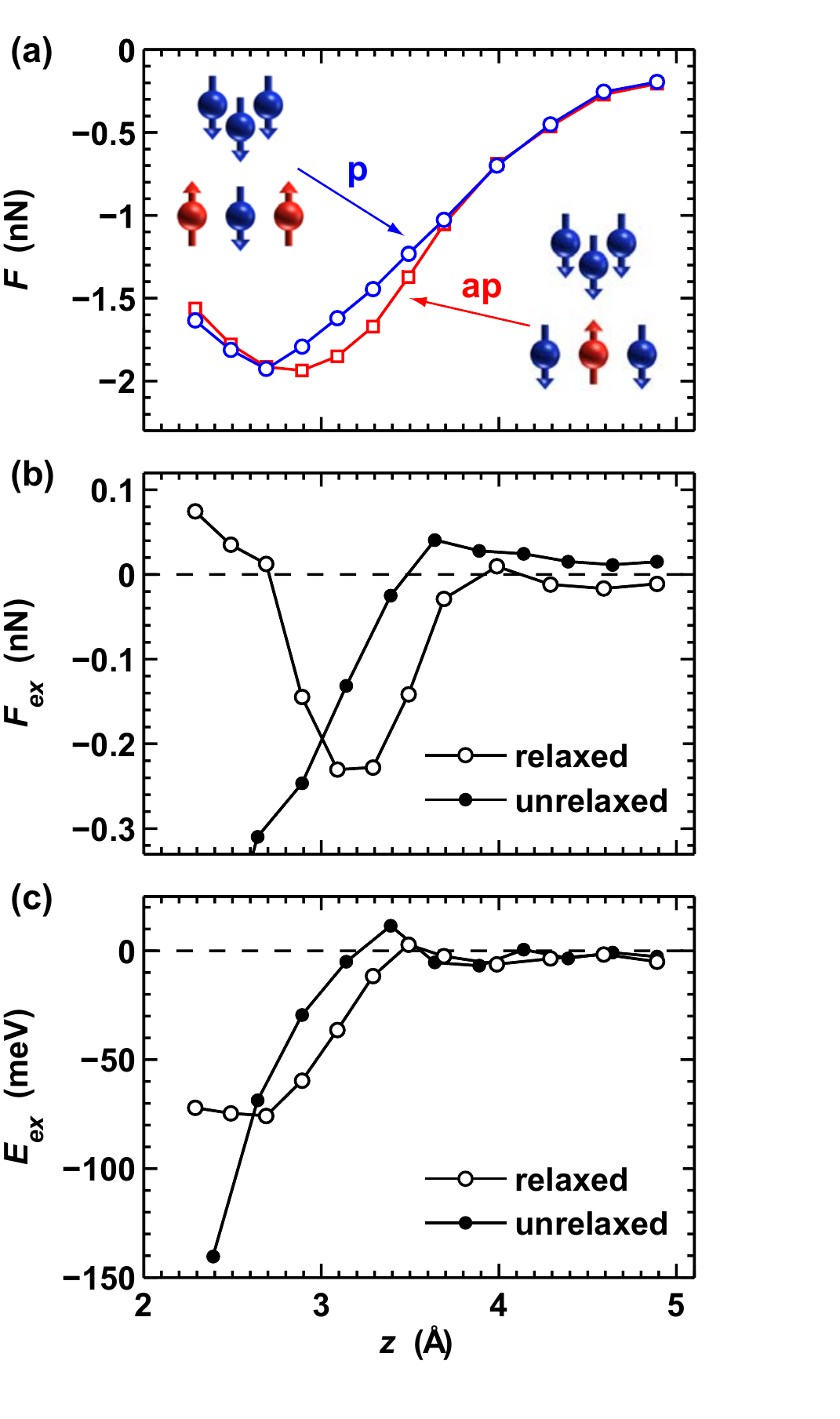}}
\caption{(Color online) \label{forcerel} (a) Calculated force
curves for the five-atoms Fe tip as it approaches the Fe ML on
W(001) on the ap-site, $F_{\text{ap}}(z)$, and on the p-site,
$F_{\text{p}}(z)$, including relaxations of tip and sample due to
their interaction. (b) Comparison of the magnetic exchange forces,
$F_{\text{ex}}(z)=F_{\text{ap}}(z)-F_{\text{p}}(z)$, and (c)
magnetic exchange energies,
$E_{\text{ex}}(z)=E_{\text{ap}}(z)-E_{\text{p}}(z)$, between the
calculations with and without relaxations as a function of the
separation between the unrelaxed tip apex and probed Fe surface
atom.}
\end{figure}

We can rationalize these force curves by assuming ferromagnetic
exchange coupling between individual Fe atoms of tip and sample
and summing over the pair-wise exchange interactions. At large
tip-sample separations, the distance between the apex atom and the
Fe surface atom beneath it is not much smaller than the distance
of the apex atom to the four neighboring Fe surface atoms.
Therefore, the ferromagnetic coupling of the Fe apex atom with the
four neighboring surface Fe atoms can dominate over the
interaction with the single Fe atom beneath the tip apex. An
antiparallel alignment with respect to the Fe surface atom beneath
the tip is then favorable. This situation results in a negative
exchange force, c.f.~Eq.~(\ref{eq:MExF}), which increases with
decreasing distance due to the larger wave function overlap. At
small tip-sample separations, however, the direct ferromagnetic
coupling of the apex atom with the Fe surface atom beneath it
becomes large and a parallel alignment is preferred. This leads to
a positive contribution to the exchange force on the apex and the
decrease of the exchange force at small separations. Of course,
this simple discussion neglects that the exchange interaction
between individual Fe atoms has a distance dependence of its own.
In addition, the magnetic moments of the Fe atoms are not constant
upon the approach of the tip as we will see in section
\ref{sec:interaction}.

\subsection{Influence of structural relaxations}
\label{sec:relaxations}

The calculations without structural relaxations presented in the
last section for the five-atoms tip showed that significant forces
act on the tip apex atom depending on the magnetic configuration
between tip and sample. From these results we conclude that
relaxations of tip and sample due to the magnetic interactions can
play an important role for the total detectable exchange force.
Therefore, we carried out the same set of calculations as before
but this time we performed a structural relaxation of the tip apex
atom and the first two layers of the sample at every tip-sample
separation. Since the tip apex atom is allowed to relax, the
detectable force acting on the entire tip is given by the
$z$-component of the force on the tip base atoms.

The obtained forces acting on the tip are shown in
Fig.~\ref{forcerel}(a) for the calculation including relaxations
due to tip-sample interactions. They look qualitatively similar as
the forces for the unrelaxed structure, c.f.~Fig.~\ref{force}(a);
however, quantitative differences arise in their respective
exchange forces which can easily be observed in the splitting
between the force curves on the p- and ap-site. Upon including
relaxations, the onset of large magnetic exchange forces shifts
towards larger tip-sample distances as seen in
Fig.~\ref{forcerel}(b). This effect facilitates their experimental
detection as the atomic force microscope can be operated at larger
distances from the point where a snap-to-contact can occur. In
addition, $F_{\text{ex}}(z)$ for the relaxed case does not display
a marked change of sign at large tip-sample distances. Similar
differences are also observed in the exchange energy for the
relaxed and unrelaxed cases, Fig.~\ref{forcerel}(c). Still,
antiferromagnetic alignment ($E_{\text{ex}}<0$) of the Fe tip with
respect to the probed Fe surface atom is energetically much more
favorable at small separations. As explained in the previous
section, the tip apex atom interacts not only with the probed Fe
surface atom but also with the four neighboring Fe atoms in the
surface with antiparallel magnetic moments. Therefore, the
negative exchange energy does not exclude ferromagnetic exchange
coupling between the magnetic moments of individual Fe atoms.

\begin{figure}
\centerline{\includegraphics[width=0.42\textwidth,viewport=0 150 300 550]{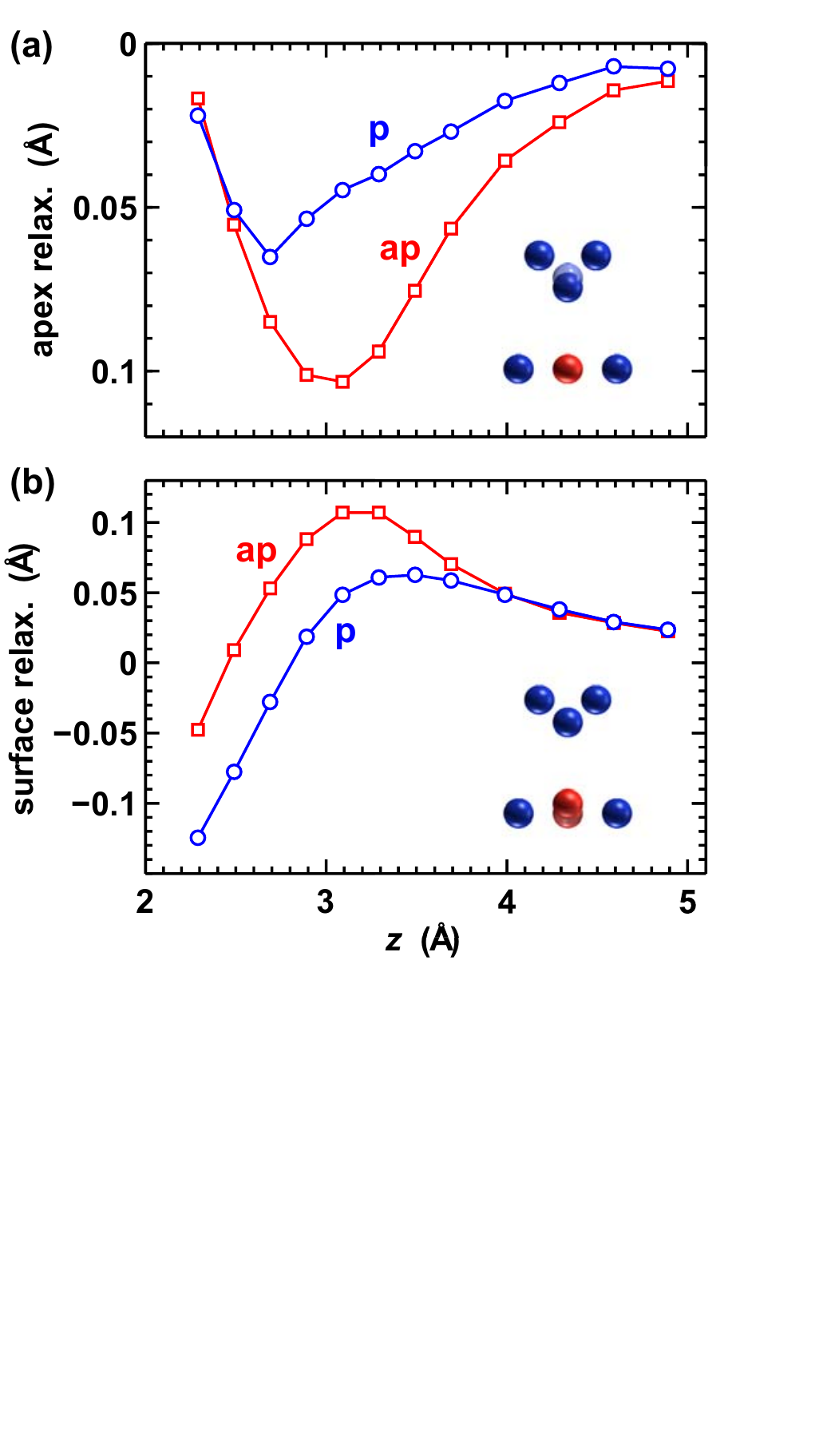}}
\caption{(Color online) \label{relax} Distance dependence of (a)
the vertical tip apex atom relaxation for the five-atoms tip and
(b) the vertical relaxation of the probed surface Fe atom for the
p- and ap-alignment, respectively.}
\end{figure}

These differences in the exchange forces and energies are
obviously caused by the relaxation of the tip apex atom which
depends sensitively on its local magnetic configuration with
respect to the approached Fe surface atom (see
Fig.~\ref{relax}(a)). The tip apex atom relaxes towards the
surface due to the attractive forces and the shape of the
relaxation curve, Fig.~\ref{relax}(a), resembles the force curves
of the apex atom, c.f.~Fig.~\ref{ForcesDecom}(a). A similar effect
is observed for the relaxation of the surface atom being probed
which is attracted towards the tip at large distances and repelled
at very close separations (see Fig.~\ref{relax}(b)). On the
ap-site, the tip apex atom relaxes about 0.05~\AA\ closer towards
the surfaces atom than on the p-site which enhances the exchange
interaction which can be inferred from the force curves of
Fig.~\ref{ForcesDecom}(a).

\subsection{Electronic and magnetic structure changes due to tip-sample interaction}
\label{sec:interaction}

\begin{figure}
\centerline{\includegraphics[width=0.42\textwidth,viewport=0 150 300 550]{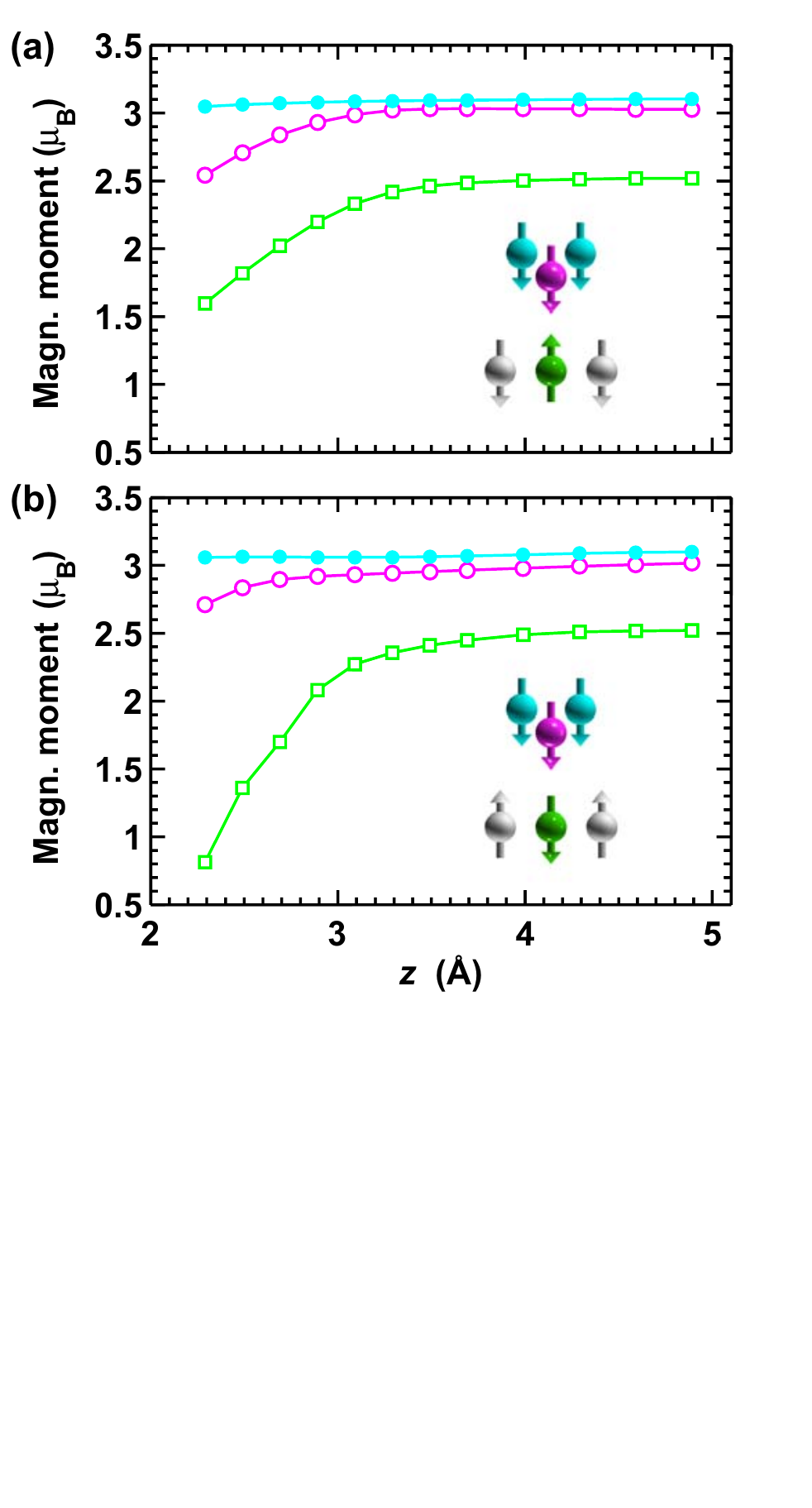}}
\caption{(Color online) \label{moment} Distance dependence of the
absolute magnetic moments of the apex (open circles), base (close circles), and surface (open squares) Fe atoms
in the case of the five-atoms Fe tip for (a) ap- and (b)
p-alignment between the magnetization of tip and surface atom as
shown in the insets including structural relaxations of tip and
sample.}
\end{figure}

\begin{figure*}
{\includegraphics[width=0.8\textwidth]{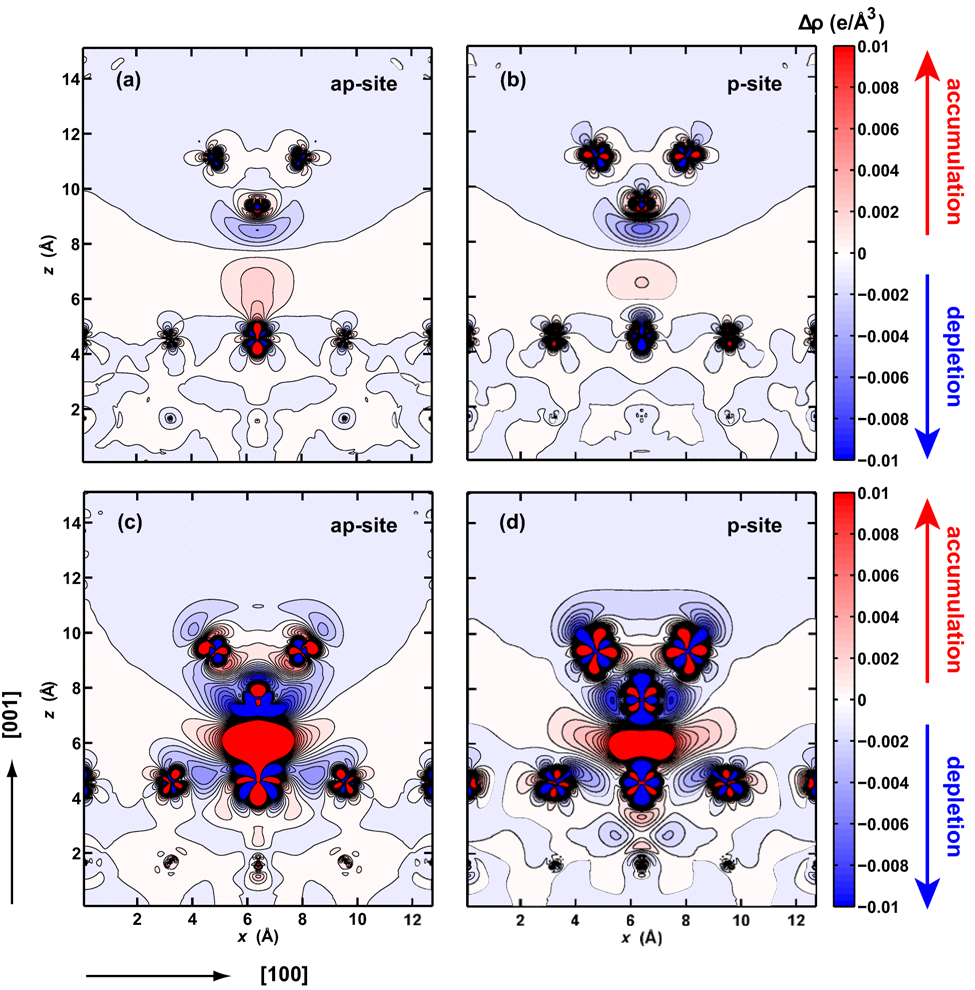}}
\caption{(Color online) \label{CDiff} Cross-sectional charge
density difference plots along the [011]-direction for the
interaction of the five-atoms Fe tip (top of each panel) with the
Fe monolayer on W(001) (bottom of each panel) at tip-sample
separations of $z=4.9$~\AA\ for (a) the ap- and (b) the
p-alignment and at $z=2.9$~\AA\ for (c) the ap- and (d) the
p-alignment. Zones in red and blue denote charge accumulation and
depletion, respectively. The results presented here correspond to
geometries including relaxations due to tip-sample interaction.}
\end{figure*}

After analyzing the interaction between tip and sample based on
force curves and the resulting relaxations in the previous
sections, we now turn to the modifications of the electronic and
magnetic structure due to their interaction. One way to monitor
the magnetic interaction is to plot the distance dependence of the
magnetic moments of tip apex atom and surface atom. This is shown
in Fig.~\ref{moment} for the two different magnetic configurations
including structural relaxations.

We find that the magnetic moment of the base atoms remains
constant at $m_{\rm base}\approx$ 3 $\mu_B$, whereas the apex and
surface atom moments decrease as the tip approaches the surface.
This decrease becomes significant only at separations below
3~{\AA} and is due to an increased hybridization between the
states of tip apex atom and surface atom. The magnetic moment drop
is more pronounced on the p-site than on the ap-site. This result
is consistent with the ap-configuration (antiferromagnetic
coupling) being energetically much more favorable than the
p-configuration (Fig.~\ref{forcerel}(c)), as there is a large
energy cost to reduce the magnetic moments from their equilibrium
values (obtained at large tip-sample separations).

\begin{figure}
\centerline{\includegraphics[width=0.45\textwidth]{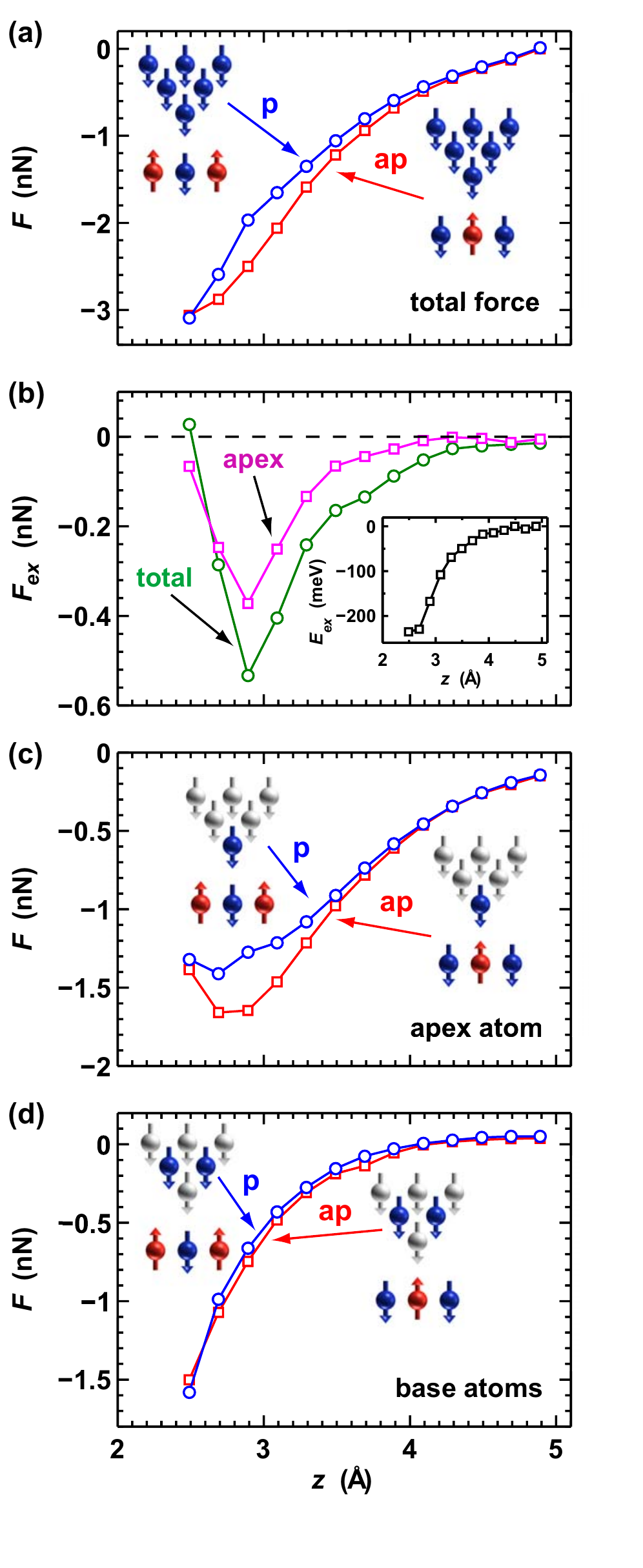}}
\caption{(Color online) \label{Forces_14atoms_tip} (a) Calculated
force curves between an Fe tip consisting of 14 atoms,
c.f.~Fig.~\ref{models}, and an Fe ML on W(001) for a parallel (p)
and antiparallel (ap) alignment of the tip magnetization and the
probed Fe surface atom. Structural relaxations due to tip-sample
interaction have been neglected in this case. (b) total exchange
force on the tip and exchange force on the apex atom. Inset shows
the exchange energy. (c) forces acting on the apex atom of the
tip. (d) forces acting on the base atoms of the tip.}
\end{figure}

\begin{figure}
\centerline{\includegraphics[width=0.45\textwidth,viewport=0 350 300 550]{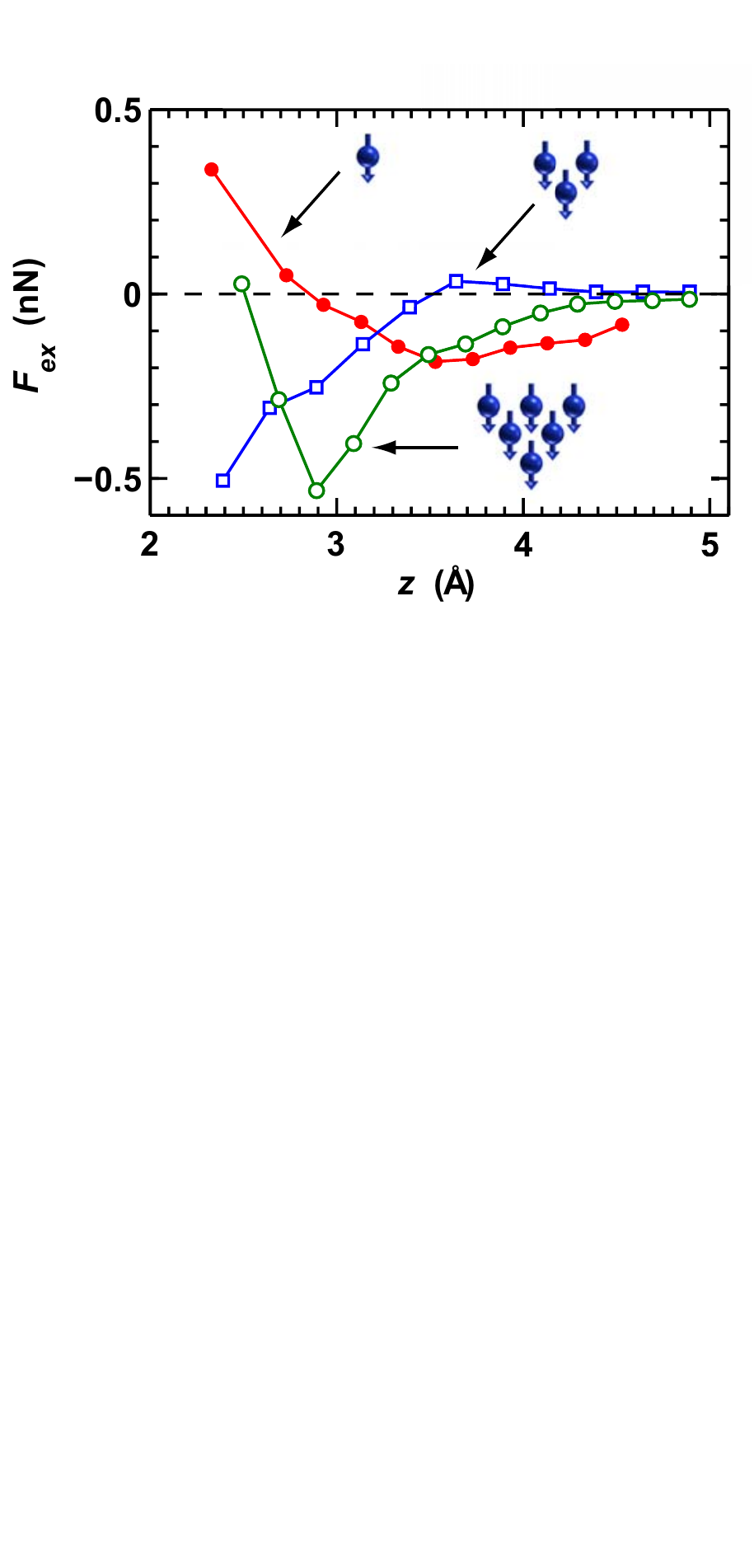}}
\caption{\label{Fex_vs_tip_size} (Color online) Comparison of the magnetic
exchange force $F_{\rm ex}(z)$ as a function of tip-sample
separation $z$ on the Fe monolayer on W(001) using a single Fe
atom tip, a five Fe atoms tip, and an Fe tip consisting of 14
atoms, as shown in Fig.~\ref{models}. All force curves presented
in this plot have been obtained without structural relaxations of
tip and sample due to their interaction.}
\end{figure}

The origin of the magnetic exchange interaction can be traced to
the different electronic interactions in the ap- and
p-configuration. In order to study the nature of these
interactions it is helpful to analyze \emph{charge density
difference} (CDD) plots for the two types of coupling. This
quantity is obtained by subtracting from the charge density of the
interacting system consisting of Fe cluster tip and 1ML Fe/W(001)
both the charge density of the isolated Fe ML/W(001) and that of
the isolated Fe cluster tip, using the relaxed atom positions also
for the isolated systems. The CDD plots allow the visualization of
the charge transfer associated with the electronic interaction
between tip and sample, i.e.~the accumulation or depletion of
charge.

Figure \ref{CDiff} shows the CDD plots for the ap- and p-alignment
at tip-sample distances of $z=4.9$ and $2.9$~\AA. At the large
separation, there is a small net charge accumulation between the
tip apex atom and the surface Fe atom. Already at this height the
interaction depends on the type of spin alignment. The charge
accumulation due to tip-sample interaction in the ap-configuration
is bound to the Fe surface atom and has a node with the Fe apex
tip atom, while in the p-configuration, it has nodes on both the
Fe surface and the tip apex atom. At a very close distance of
$z=2.9$ \AA\, electronic charge strongly accumulates between the
tip apex atom and the surface Fe atoms, implying a strong
electronic interaction between the tip and the surface. The charge
accumulation in the ap-coupling is larger than in the p-coupling
in agreement with the ap-alignment being energetically more
favorable, c.f.~Fig.~\ref{forcerel}(c).

The CDD plots also show that the charge density of the
nearest-neighbor Fe atoms (with respect to the probed Fe surface
atom) is considerably redistributed upon approaching the tip.
Therefore, the exchange coupling of these nearest neighbor Fe
atoms with the apex atom of the tip plays an important role to
determine whether p- or ap-alignment is more favorable. Similarly,
the redistribution of the base atom's charge density indicates a
significant contribution to the exchange interaction between tip
and sample as has been discussed in terms of the exerted exchange
forces in section \ref{sec:unrelaxed},
c.f.~Fig.~\ref{ForcesDecom}.

\subsection{Influence of tip size}
\label{Cmodels}

One of the more delicate aspects in modelling atomic force
microscopy experiments is the geometry used for the tip. Ideally,
the tip should consist of thousands of atoms to mimic the tips
used in real experiments. However, in practice one is limited by
the computational resources required for the calculation.
Fortunately, the chemical and magnetic interaction between tip and
sample is dominated by the foremost atoms due to the exponential
decay of the wave functions while long-range forces can be added
based on continuum models.\cite{Giessibl1997} However, the
electronic and magnetic properties at the tip apex is still
influenced by the base of the tip used in the model and needs to
be investigated.

In the past, theoretical calculations have often been carried out
using a single Fe atom as an idealized model of the tip to study
the magnetic exchange force e.g.~on the NiO(001)
surface.~\cite{2005SurSc.590...42M} Here, we assess the validity of
such a model using the Fe monolayer on W(001) as a test sample by
comparing calculations with a single Fe atom with the five Fe
atoms pyramid tip discussed in the previous sections and an even
larger fourteen Fe atoms tip, c.f.~Fig.~\ref{models}.

Fig.~\ref{Forces_14atoms_tip}(a) displays the calculated
force-distance curves for the interaction between the 14 atoms Fe
tip and the Fe ML on W(001). Because of the large tip size and the
resulting prohibitive computational effort we neglected structural
relaxations due to the tip-sample interaction in this case.
Similar to the calculation with the five Fe atoms tip,
c.f.~Fig.~\ref{force}, we observe a splitting between the total
forces acting on the tip for the p- and ap-site which increases
with decreasing tip-sample separation as shown in
Fig.~\ref{Forces_14atoms_tip}(a). The resulting exchange force and
energy displayed in Fig.~\ref{Forces_14atoms_tip}(b) are negative
which indicates a preferred antiferromagnetic coupling with the
probed surface Fe atom. This exchange interaction is in good
qualitative agreement with the results we obtained for the smaller
five atoms Fe tip, c.f.~Fig.~\ref{force}.

\begin{figure}
\centerline{\includegraphics[width=0.40\textwidth,viewport=0 150 300 550]{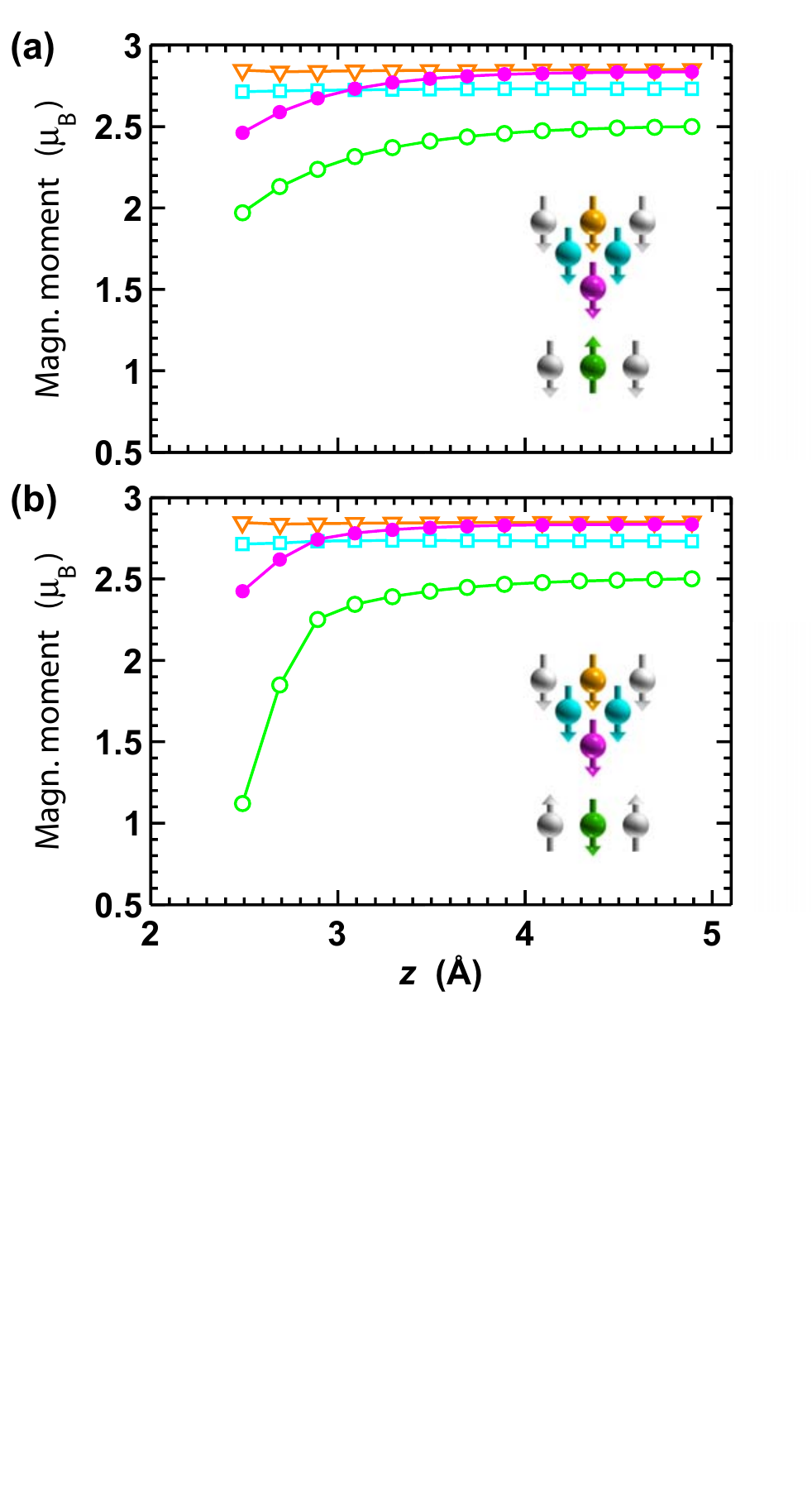}}
\caption{\label{Moments_14atoms_tip} (Color online) Distance dependence of the
magnetic moments of different Fe atoms of the 14 atoms tip -- apex (closed circles), base second layer (open squares), center atom of third layer (open triangles) -- and the
surface Fe atom (open circles) of the Fe ML on W(001) for (a) ap- and (b)
p-alignment between the magnetization of tip and surface atom as
shown in the insets without structural relaxations of tip and
sample. }
\end{figure}

However, the decomposition of the forces on the distinct tip
atoms, Figs.~\ref{Forces_14atoms_tip}(c) and (d), displays
interesting discrepancies between the two tip models. The forces
acting only on the tip apex atom are qualitatively quite similar,
e.g.~the forces are larger for ap-configuration, however, the
splitting between the two curves is larger for the five atoms tip.
For the base atoms, there is a more dramatic difference. Here, we
obtain a smaller exchange force for the base atoms in the larger
tip and the sign of the exchange force on the base atoms is the
same as for the apex atom. However, the contribution of all base
atoms to the total exchange force is still significant, as seen
from $F_{\rm ex}^{\rm apex}(z)$ given in
Fig.~\ref{Forces_14atoms_tip}(b). For the smaller tip, the sign of
the exchange force on the base atoms was opposite to that of the
apex atom thereby reducing the total exchange force. For the 14
atoms tip their sign is the same, and consequently, the regime of
considerable exchange forces sets in at larger tip-sample
separations as seen in Fig.~\ref{Fex_vs_tip_size}.

\begin{figure*}
{\includegraphics*[width=0.8\textwidth]{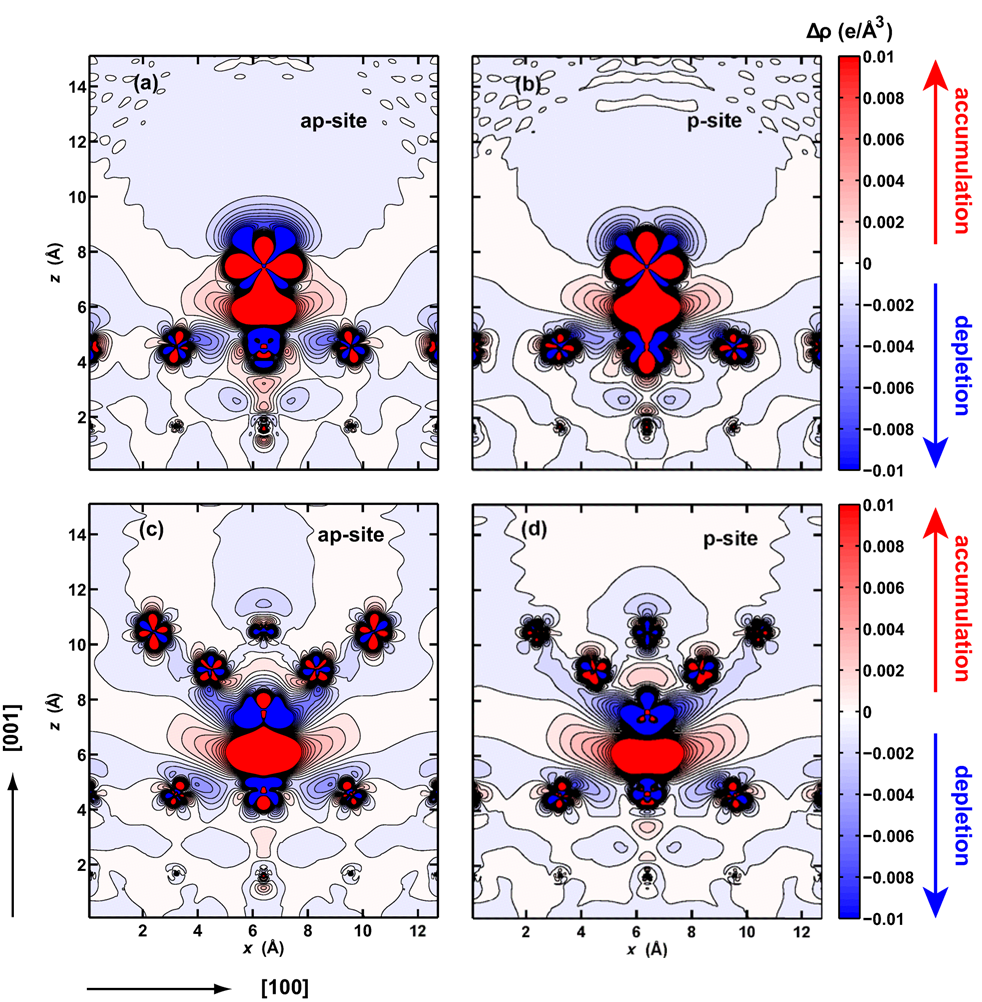}}
\caption{\label{BigTipEDD} (Color online) Cross-sectional charge density
difference plots along the [011]-direction for the interaction of
the single Fe atom tip (top of each panel) with the Fe monolayer
on W(001) (bottom of each panel) at a tip-sample separation of $z=
2.9$ \AA\ for (a) the ap-coupling and (b) the p- coupling. (c) and
(d) show equivalent plots for the Fe tip consisting of 14 atoms.
Zones in red and blue denote charge accumulation and depletion,
respectively. The results presented here correspond to unrelaxed
geometries for the single Fe atom tip and relaxed geometries for
the 14 atoms tip.}
\end{figure*}

A direct comparison of the exchange forces for different tip
models is given in Fig.~\ref{Fex_vs_tip_size}. Obviously, the
exchange force obtained for a single Fe atom tip is even
qualitatively different from both pyramid-type tips. At large
separations, the exchange forces are much larger than for the
pyramid tips, while they have the opposite sign at close distance.
From these calculations, it is quite clear that a single Fe atom
cannot mimic the exchange forces between a magnetic tip and
sample. If we compare the two pyramid-type Fe tips, on the other
hand, the general shape of the curve is very similar and the
smaller tip gives qualitatively the same result. However, the
exchange forces for the bigger Fe tip are significantly enhanced
and set in at much larger tip-sample distances which is of crucial
importance in experiments.

The dependence of the tip's magnetic moments on the tip-sample
separation, displayed in Fig.~\ref{Moments_14atoms_tip} for the 14
atoms Fe tip, provides additional insight into the tip size
dependent magnetic interaction. Similar to the smaller five-atoms
tip, only the magnetic moment of the foremost tip atom is reduced
upon the approach to the surface. However, the moment of the apex
atom is smaller, c.f.~Fig.~\ref{models}, and its relative
reduction due to tip-sample interaction is slightly enhanced for
the larger tip. In contrast, the magnetic moment of a single Fe
atom tip is practically constant upon approaching the surface (not
shown) and the exchange force is smaller,
c.f.~Fig.~\ref{Fex_vs_tip_size}. Therefore, we conclude that a
large magnetic moment of the apex atom does not guarantee
significant magnetic exchange forces. Instead, a less rigid
magnetic moment of the apex atom, i.e.~tunable in size due to the
interaction, is favorable to detect large exchange forces.
In addition, the exchange forces on the base atoms of the tip play
an important role for the total exchange force as discussed above.

In order to check the influence of using the larger 14 atoms
pyramid tip on the relaxations, we performed a structural
optimization for a single tip-sample distance of $z=2.9$ \AA. As
for the five atoms tip, we find that due to the attractive
interaction the apex atom relaxes towards the surface. The
relaxation values for the 5 Fe atoms cluster tip are 0.11 and
0.05~\AA\ for the ap- and p-alignment, respectively, while for the
14 atoms tips the values are 0.16 and 0.12~\AA\ for the ap- and
p-alignment, respectively. The difference in the relaxations are
in nice qualitative agreement for the 5 and 14 atoms tips, in
particular, the apex atom relaxes more in the ap- than in the
p-alignment. At this close tip-sample separation of 2.9~{\AA}, the
total exchange force on the 14 atoms tip after relaxation is
nearly unchanged while it actually decreases for the five-atoms
tip, c.f.~Fig.~\ref{forcerel}(b).

Further evidence for the modified tip-sample interaction of a
single-atom tip and multi-atom tips can be obtained by examining
the CDD plots. These are shown in Fig.~\ref{BigTipEDD} for the
single Fe atom tip and the 14 Fe atoms tip at a separation of
$z=2.9$~\AA\ which can be directly compared to those for the 5
atoms tip, displayed in Figs.~\ref{CDiff}(c) and (d). These graphs
show a similar redistribution of electronic charge density between
tip apex and surface atoms for the two multi-atom tips upon
approaching the tip to the surface. The dependence on the two
types of spin alignment is also quite similar. Interestingly, the
charge redistribution of the base atoms does not depend as
dramatically on the type of coupling for the 14 atoms tip as for
the 5 atoms tip. This can explain the smaller exchange forces
acting on the base atoms for the larger tip. In contrast, the
electronic charge distribution for the single atom tip, shown in
Figs.~\ref{BigTipEDD} (a) and (b), is distinctively different from
the case of the pyramid-type tips. In fact, the CDD plots for the
ap- and p-alignment of the single-atom tip are very similar, which
explains the very small exchange force at $z=2.9$~\AA\ for the
single atom tip observed in Fig.~\ref{Fex_vs_tip_size}. Therefore,
one should use multi-atom tips in {\em ab initio} simulations of
magnetic exchange force microscopy in order to properly describe
the electronic and magnetic properties of the tip as well as the
magnetic interaction with the sample.

\subsection{Implications for spin-polarized STM}
\label{sec:spstm}

So far, we have interpreted our calculations on the magnetic
tip-sample interaction only with respect to magnetic exchange
force microscopy. However, a similar situation occurs in a
spin-polarized STM experiment which relies on measuring the
spin-polarized tunneling current between a magnetic tip and a
magnetic sample.

In the constant-current mode the current is fixed while scanning
the tip across the surface by approaching or retracting the tip in
the vertical direction. Due to the variation of the spin-polarized
local density of states in the vacuum the constant-current mode
allows to resolve magnetic structures on the atomic
scale,~\cite{2000Sci...288.1805H,PhysRevLett.86.4132} e.g.\ the
antiferromagnetic order of the Fe monolayer on
W(001).~\cite{2005PhRvL..94h7204K} In simulations of STM
experiments, one often neglects structural relaxations of the tip
while it is scanned across the sample. However, in some cases they
can lead to large enhancements of the corrugation
amplitude,~\cite{DiVentra1999,Hofer2001,PhysRevB.70.085405}
i.e.~the maximum vertical variation of tip position while it is
scanned across the surface.

\begin{figure}
{\includegraphics[width=0.40\textwidth]{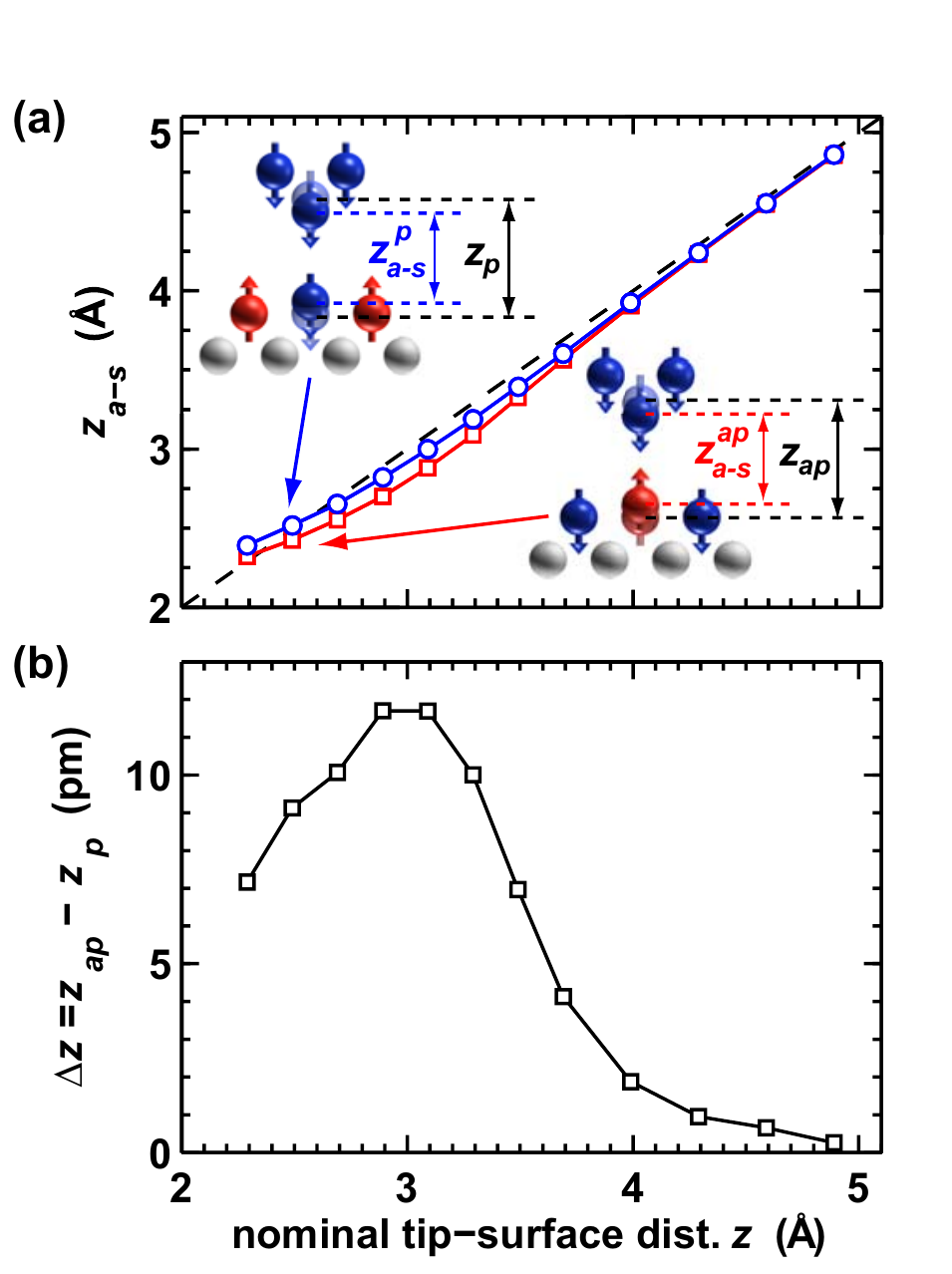}}
\caption{\label{corrug} (Color online) (a) Actual separation between the tip apex
atom of the five Fe-atoms tip and the surface atom of the Fe
monolayer on W(001) including structural relaxations, $z_{\rm
a-s}$, as a function of the separation before relaxation, $z$,
c.f.~inset, for the p- and ap-alignment between the magnetic
moments of tip apex and surface atom. Dashed line indicates the
unrelaxed case. (b) Effective corrugation amplitude which would
occur in a spin-polarized STM experiment due to the magnetic
configuration dependent tip and sample relaxations.}
\end{figure}

As we have seen in section \ref{sec:relaxations}, tip relaxations
can depend on the local magnetic configuration between tip and
sample magnetization. As the tunneling current depends
exponentially on the tip-sample separation and is dominated by the
foremost tip atom, a relaxation of the apex atom can drastically
change the measurable corrugation amplitude. In a simple model, we
can assume that the tunneling current depends exponentially on the
actual distance between the tip apex atom and the surface atom
underneath it which we denote by $z_{\rm a-s}$. This distance is
given by
\begin{equation}
z_{\rm a-s} (z) = z - \Delta z_{\rm a}(z) - \Delta z_{\rm s}(z),
\end{equation}
where $z$ is the nominal separation without relaxations and
$\Delta z_{\rm a}(z)$ and $\Delta z_{\rm s}(z)$ denote the
relaxations of tip apex and surface atom, respectively. These
relaxations are displayed in Fig.~\ref{relax} for the parallel (p)
and antiparallel (ap) alignment of the magnetic moments of the
apex atom of the five-atoms Fe tip and the probed Fe surface atom
of the Fe ML on W(001). The actual separation between the apex and
the surface atom is plotted in Fig.~\ref{corrug}(a) as a function
of the nominal distance $z$ for the ap- and p-configuration. Quite
obviously, there is a significant deviation from the linear
unrelaxed case, which indicates that the relaxation are
significant. Also a clear difference between the curves for the p-
and ap-site can be observed.

In order to obtain a constant tunneling current $z_{\rm a-s}$ must
be the same on both the p- and ap-site of the surface. Therefore,
the corrugation amplitude $\Delta z = z_{\rm ap} - z_{\rm p}$ due
to structural relaxations is given by
\begin{equation}
\Delta z (z)=(\Delta z_{\rm a}^{\rm ap} (z) + \Delta z_{\rm
s}^{\rm ap}(z)) - (\Delta z_{\rm a}^{\rm p}(z) + \Delta z_{\rm
s}^{\rm p}(z)),
\end{equation}
where the upper index denotes the relaxations on the ap- and
p-site, respectively. This quantity is shown in
Fig.~\ref{corrug}(b) as a function of the unrelaxed apex-surface
distance $z$ between the Fe tip and the Fe ML on W(001). There is
a very steep rise of this apparent corrugation amplitude below a
nominal tip-sample distance of 4 \AA\ at which it is 2~pm, while
beyond this distances, the effect of structural relaxations
becomes very small. The SP-STM experiments on the Fe monolayer on
W(001)~\cite{2005PhRvL..94h7204K,2006NatMa...5..477B} reported
corrugation amplitudes between 3 and 10~pm. The absolute
tip-sample distance is unknown in STM, however, in some cases the
tunneling parameters, i.e.~bias voltage and tunneling current,
hint at small separations. Therefore, our results indicate that
contributions due to structural relaxations of the tip may play a
role in some SP-STM measurements.

In our discussion, we have so far neglected that the
spin-polarized tunneling current is different on the two magnetic
surface sites of the Fe ML on W(001). This contribution to the
current depends on the spin-polarization of the local density of
states of tip and sample close to the Fermi energy. In the
constant-current mode it causes a different tip-sample separation
on the two Fe surface atoms of opposite spin direction and allows
the resolution of the atomic-scale magnetic
structure.~\cite{2005PhRvL..94h7204K} For the total corrugation
amplitude both effects, i.e.~spin-polarization of the tunneling
current and spin-alignment dependent relaxations, are additive.
However, the two contributions can be of opposite sign as the
first depends on the spin-polarization at the Fermi energy while
the later stems from all occupied states. Therefore, the
corrugation amplitude due to structural relaxations can either
enhance or diminish the corrugation from the spin-polarized
current.

\section{Summary}
\label{sec:conclusion}

We have performed first-principles calculations based on density
functional theory to study the interaction between a magnetic tip
and a magnetic sample which occurs in magnetic exchange force
microscopy (MExFM). We have studied tips consisting of one to
fourteen Fe atoms and have chosen one monolayer Fe on W(001) as a
sample system which exhibits an antiferromagnetic checkerboard
structure and has been resolved on the atomic-scale by both
MExFM~\cite{Schmidt2008} and spin-polarized scanning tunneling
microscopy (SP-STM).~\cite{2005PhRvL..94h7204K}

Our calculations clearly demonstrate the inadequacy of using a
single magnetic atom as a model for the magnetic tip in MExFM as
the obtained force curves differ even qualitatively from those of
cluster tips. Increasing the size of our Fe bcc(001)-type pyramid
tip still leads to quantitative changes, however, qualitatively
the five and fourteen Fe atoms tips exhibit the same features in
the force-distance curves and lead to antiferromagnetic exchange
coupling with the probed Fe surface atom being energetically more
stable. Quantitatively, we observe that the onset of significant
exchange forces is shifted to larger tip-sample separations for
the larger tip.

The exchange forces on the apex atom is the dominant contribution
to the total exchange force for both tips, but contributions from
other tip atoms cannot be neglected and may even reduce the total
exchange force. This effect is especially pronounced for small
tips. The chemical and magnetic interaction of the Fe apex atom is
significant with both the Fe surface atom underneath it as well as
the nearest-neighbor Fe surface atoms.

We find that structural relaxations of tip and sample due to their
chemical and magnetic interaction play an important role and can
greatly enhance the measurable MExFM signal. These relaxations
depend on the local magnetic configuration of tip and sample
magnetization and can have an influence on SP-STM experiments as
well. In particular, the effective corrugation amplitude of the
magnetic superstructure observable in SP-STM can be enhanced or
diminished.

\begin{acknowledgments}
It is our pleasure to thank S.~Bl\"ugel, Y.~Mokrousov,
P.~Ferriani, A.~Schwarz, U.~Kaiser, R.~Schmidt, and
R.~Wiesendanger for many insightful discussions. Computations were
performed at the Hamburg University of Technology, the
Norddeutscher Verbund f\"{u}r Hoch- und H\"{o}chstleistungsrechnen
(HLRN), and the Forschungszentrum J\"ulich (JUMP). We acknowledge
financial support from the DFG (Grants No.\ HO 2237/3-1 and HE
3292/4-1). S.H.\,thanks the Stifterverband f\"ur die Deutsche
Wissenschaft and the Interdisciplinary Nanoscience Center Hamburg
for financial support.
\end{acknowledgments}


\end{document}